\documentstyle{article}

\begin{document}
\title{
The developing structure of dynamical systems
}

\date{}
\author{
H. P. Fang
\\{\small Department of Physics, Fudan University,
        Shanghai 200433, China;$^*$}
        \\{\small Institute of Theoretical Physics, P.O. Box. 2735,
        Beijing, 100080, China.}
	}
\vskip.15in
\maketitle
\begin{abstract}
 In order to investigate the evolutionary process of many deterministic
dynamical systems with unfixed parameter, a set of dynamical models with
parameter changing continuously and the accumulation of this
change might be large is introduced and discussed. The boundary crises and
the period-doubling bifurcations are found suppressed and scaling
properties for these phenomena are exploited. Due to this suppression,
the period-doubling bifurcations seem no longer continuous.
We further discuss the
possible applications of these models.

\end{abstract}
\vskip.15in

$Introduction$.
One of the practical methods to investigate the phenomena common
in many scientific fields such as physics, biology, ecology,
 geophysics and economy, etc is by modeling them either with differential,
difference-differential, or pure difference equations.
After the models have been studied, these phenomena can be
understood and even be predicted. In this Letter,  we will introduce
a model which can be used to model and then study the behavior of many
natural systems with unfixed environment parameters.

The motivation of introducing such models is our recent carefully
examination on the work  which was tempted to apply the chaotic dynamics
to understand and predict the natural phenomena [see, e.g. ref 1-5].
C. Nicolis and G. Nicolis proposed some
evidence for the climate which caused a debate on whether climatic
attractors exist or not$^{1}$. The correlation dimensions$^6$ for
many economical data sequences had also been calculated which were taken as
the evidence that the economical chaos might exist [see, e.g., ref. 2].
And recently, the short term prediction algorithm was used to test
the predictability   of some economical data$^3$. Chaos in heart and brain
was also reported and tempted to be controlled$^{4}$. All of these
studies are based on the assumption that these natural systems are
environment independence, or the data of these dynamical systems are
take in a limited time interval so that accumulation of the environment
parameter change is sufficiently small. Then
the familiar concepts such as
phase-space, bifurcation diagrams, orbits, attractors, return maps, basins
of attractor, manifolds, etc. can be used to extract information
 of these natural systems $^{7-8}$.
In fact, many of the  natural systems such as all biological, ecological
 and economical systems have developed into other ones before they reach
any asymptotic states.
The environment parameters change so quickly that accumulation of the
parameter change is {\bf significantly large}
in a  limited time interval which has changed the dynamical behavior
of the systems.
The data sequences from these systems, on one hand, are some kinds
of transients in short term interval (so that the accumulation
of the parameter change is small), and on the other hand, are
the sets of points in different environment parameters in a relative
long time interval.
 It is clear that the above listed techniques are applicable
only when the change of
environment is negligible. Thus, this assumption not only
restricts the study on the global behavior of these natural systems,
but also might cause misunderstanding.

        In order to study the dynamical behavior of these natural
systems, and clarify the possibility and limitations to
apply the familiar concept to study natural systems, in this
Letter we will study the dynamical models with
$parameter\ changing\ continuously$ and $the$ $ accumulation$ $of\ this
\ change\ might\ be\ large$. Explicitly,
we will study the models in the form
\begin{equation}
\vec{X}_{n+1} =  F(\vec{X}_n, \vec{P}_n),
\end{equation}
\begin{equation}
\vec{P}_{n+1} = G(\vec{P}_n)+\vec{P}_n.
\end{equation}
where $\vec{X}_n$ and $\vec{P}_n$ represent one-, two- or
high-dimensional variable and parameter vectors, respectively.
Considering the continuous evolutionary process of the natural systems,
$G(\vec{P}_n)$ is usually small (If Eq. (1) represents a difference form
of differential equations with very small time step $\delta t$,
$G(\vec{P}_n) \rightarrow 0$ as $\delta t \rightarrow 0$).
One particular case for this model has already been discussed in detail
which is the periodic forced driven systems.
Some of the special cases of this model with the parameter close to
the bifurcation points or $G(\vec{P}_n)$ is a noise had also
received attention recently $^{9-11}$. In the present Letter,
we will concentrate on the situation with the parameters change
unperiodically and ever irregularly, and the accumulation
of the change of the parameters might be large (hereafter we call these systems
the structure-variable systems, namely, SVS's) . We will discuss the behavior
and possible applications for these models.

Due to the unperiodic change of the parameters, the dynamical systems
will never approach asymptotic states. All the data we observed are
transients. When the change of the parameters is not too fast
(G($\vec{P}_n$) is sufficiently small), for a short observation time
interval, the SVS represented by eqs. (1-2) might exhibit
some behavior similar to those of Eq. 1 with fixed parameters $\vec{P}_n$.
 For $\vec{P}_n$ close to the parameters for bifurcations or boundary
crises$^{12}$ of Eq. 1, the behavior of
the SVS (1-2) is significantly different from that
of system (1) with fixed parameters.
The crises and period-doubling bifurcations
are suppressed and  scaling properties about these phenomena
can be found. Due to this suppression
the periodic orbits and period-doubling bifurcations are difficult
to be observed directly.

$Example$ For a detail idea of the behavior of these SVS's
we take the following map
\begin{equation}
\cases{x_{n+1} =  f(\mu_n, x_n) = 1 - \mu_n x_n^2,
\cr \mu_{n+1} = \mu_n + \delta \mu,}
\end{equation}
as an example.  Fig. 1 shows a typical case for a trajectory of this map.
This diagram shares some similarity with the bifurcation diagram
of the Logistic map
in the same parameter interval. Unlike that in the bifurcation diagram,
no point is thrown away as transients. In fact, this diagram
reflects the  evolutionary process of the dynamical systems described
by the Logistic map (we will call it the developing diagram hereafter),
which corresponds to an idea model of the insects
population on an island$^{13}$ with the environment of the island changes
 (The area might shrink because of the housegreen or other effects).
 The value of the parameter step $\delta \mu$ reflects
the rapidity of evolutionary process of the dynamical system.
It is clear that when the width  of any periodic
window is small or comparable with  the value of $\delta \mu$, this
periodic window disappears in the developing diagram.

A direct observation of the Fig. 1 is that the period-doubling
bifurcation seems no longer a continuous one. Comparing to the
bifurcation diagram, the parameter of period-doubling is shifted.
Now we estimate the rapidity of the convergence of the period-doubling
bifurcations and the shift with respect to $\delta \mu$ in these
developing diagram.

Consider a period-doubling bifurcation occurs at $\mu_c$ in the bifurcation
diagram and at $\mu_d (\delta \mu)$ in the developing diagram with a parameter
step $\delta \mu$. $\mu$ is the controlling parameter. $\mu_d (\delta \mu)
> \mu_c$. In fact, from the parameter
$\mu = \mu_c$, the developing diagram goes alone the lose-stability periodic
orbit until it reaches $\mu = \mu_d$, due to the slowing down of the
convergence and the  inertia.

Consider a period $p$ orbit which should be a certain fixed point  $x^*$ of the
following map
 \begin{equation}
x_{n+p} = \underbrace{f \circ f \circ \cdots \circ f(\mu, x_n)}_{p\ {\rm
times}}.
\end{equation}
Let
$$ x_n = x^* + \epsilon_n,$$
$|\epsilon_n|$ diminishes as $e^{-A|\mu - \mu_c|}$ for each iterate of the
map (4) at parameter $\mu$, and $A$ is a positive constant$^{14}$.
 Consider that we go forward $k$ steps from $\mu =\mu_c$ of map (4) and
reach $\mu_d =\mu_c + k \delta \mu$. Assume that the
initial deviation of the period $p$ orbit is $\epsilon^{(0)}$, and the
final deviation is $\epsilon^{(k)}$,
$$|\epsilon^{(k)}| \propto \displaystyle e^{-A \sum_{i=1}^k |\mu_i - \mu_c|}
|\epsilon^{(0)}| \\
=\displaystyle e^{-A \sum_{i=1}^k i|\delta \mu|} |\epsilon^{(0)}|\\
=\displaystyle e^{-A k (k+1) |\delta \mu|/2 } |\epsilon^{(0)}|.$$
For all small $|\delta \mu|$ we assume
 $|\epsilon^{(k)}/\epsilon^{(0)}|$ to be an
approximate constant, so that $k (k+1) \delta \mu$ is approximately a constant.
Reminding that $k \delta \mu = \mu_d -\mu_c$,
 we obtain that
 \begin{equation}
|\mu_d(\delta \mu) - \mu_c| \propto \sqrt{|\delta \mu|}.
\end{equation}
We have checked this relation numerically for many period-doubling
bifurcations in many models. This
relation holds approximately provided the parameter step $\delta\mu$ is
not too large.

At $\mu = \mu_d$ the rapidity of convergence  is proportional to
$$e^{-A|\mu_d - \mu_c|}.$$  The larger the value of $|\mu_d - \mu_c|$, the
smaller
the value of $e^{-A|\mu_d - \mu_c|}$. Thus the larger the value of $\delta
\mu$,
 the more rapid the convergence (see Eq. 4). This makes the  period-doubling
 bifurcation no longer a continuous bifurcation for not too small $\delta \mu$.

The parameter for a boundary crisis is also shifted as shown in Fig. 2.
 In a similar way, we get the scaling
law for the parameter shift $\Delta \mu$ of the crisis with respect
to the controlling parameter step $\delta \mu$ as following
 \begin{equation}
\Delta \mu \propto \delta \mu/\lambda,
\end{equation}
where $\lambda$ is the maximal Lyapunov  exponent.

The parameter shifts
for both the boundary crisis and period-doubling bifurcations
cause the so-called ``artificial" $hysteresis$ that the trajectory
is still stay in a lose-stability attractor while another attractor has
already existed.

Besides the enviroment parameter changing of of many dynamical systems,
it should also pay attention on the method of extracting data from the
natural systems, in which we usually use the average or sliding average
values such as the  economical data per  hour, day, month and even year,
and the $b$-values in seismic events which are the sliding averages of
seismic events in a definite time interval$^{15,16}$. The periodic orbits
from these data are totally obscured.
In Fig. 3 we shown the sliding average values of the SVS (2).
It is clear that the periodic orbits correspond to some  dips.
This gives us a hint that a tangent bifurcation might happen at this
parameter range and periodic orbits might exist.

{\it Possible applications} There are various applications of the developing
diagram.  Recently,  short term predictability$^{17}$ on deterministic
dynamical systems with fixed parameters has been obtained a great
deal attention. Are the data from a dynamical system with changing
parameter predictable? With our model, we can show that
 that $the\ short$ $term$ $predictability$ $is$ $possible$ $for$ $ these
$ $natural$ $phenomena$ $no$ $matter$ $the$ $environment$ $parameter$ $
changes $ $slowly$ $or\ quickly$, provided that the changing of the
parameter is governed by some rules (Though we might not know exactly
what they are, see detail in our recent work, ref, 18).  With this
 idea, we can understand that it is reasonable that some economical
data$^4$ and the $b$-values$^{19}$ for seismic events are predictable.

	The developing diagram can also be used to clarify the
real physical meaning of the various characterizing quantities
(correlation dimension, Lyapunov exponent, etc) which were calculated
 for many natural dynamical systems$^{18}$. In fact, due to the changing
of parameters, the reconstructed figures for the data from a developing
diagram is very similar to that for a noised dynamical system.

 An important application of the developing diagram lies in the study of the
subsystems (or open systems) of dynamical systems.
The influence of the other system can be considered as
one or more changing parameters, which might be written
by some equations phenomenally. This can help us to
understand the restrictions of applying various perturbation
and truncation theory to nonlinear systems since a small
changing of the environment parameter(by adding high order perturbations
or truncations) will suppress or induce chaos, crises$^{20}$.

{\em Summary and discussion} In summary, we have introduced a set of
 dynamical models with
{\em parameter changing continuously and the accumulation of this
 change might be large}.  The parameters for boundary crises and
period-doubling bifurcations are shifted and scaling properties for these
phenomena are exploited. Due to these shifts, the
period-doubling bifurcations seem no longer continuous.
We also discuss other possible applications of the models.

There are still a number of interesting questions remained to be investigated.
Since the data series from our model (as well as many natural systems)
are the set of transients for different parameters, how to extend
the general methods[see, eg. 21,22] to extract information from the data series
and then control and predict them is needed. The "artificial hysteresis"
clearly has closely connection with many biological phenomena.
 The average values, also introduced
in this Letter, should be paid more attention even for that from a
dynamical systems with fixed parameters. We are afraid that  a great
part of noise we considered in many natural systems results
from the averaging process and the changing parameters.
These questions and the other applications
of the developing diagrams are now undertaken by the author and his
collaborations. Some of them will be presented in an extended paper.

$Acknowledgement$. The author sincerely  thanks Prof. Hao Bai-lin for his
encouragement, and  Prof. Chen Jinbiao, Li Qianling for stimulated
discussion and providing the data of seismic events in Tanshan region,
He also thanks Prof. Ding Ejiang,  Dr. Cao Liangyue Liu Jiuxie
and Wu Zuobing for helpful discussions,  Dr. Gh. Ardelean
for reading the manuscript.
This work was partially supported by a grant from CNSF.
\section*{}
\begin{description}
\item[{*}]{ Address for correspondence}.
\item[{[1]}]{ C. Nicolis and G. Nicolis,  Nature  {\bf 311}, 529 (1984);
{\it ibit} {\bf 329}, 523 (1987);
P. Grassberger,  {\it ibit}  {\bf 323}, 609 (1986);
{\it ibit} {\bf 329}, 524 (1987)}.
\item[{[2]}]{P. Chen, System Dynamics Review, {\bf 4}, 605(1988).}
\item[{[3]}]{ L.-Y. Cao {\it et al.}, submitted to Comput. Economics. In this
paper, several Chinese macroeconomic data sequences, e.g., the national
financial expenditure, total value of retain sales, gross output value of
industry and monetary supply, are found short-term predictable.}
\item[{[4]}]{ A. Garfinkel {\it et. al.}, Science {\bf 257}, 1230 (1992);
Steven J. Schiff {\it et. al},  Nature  {\bf 370}, 615 (1994)}.
\item[{[5]}]{G.P. Pavlos {\it et al.}, Int. J. Bif. \& Chaos, {\bf 4},
87(1994).}
\item[{[6]}]{ P. Grassberger and I. Procaccia,
Physica {\bf 9}D, 189 (1983)}.
\item[{[7]}]{ Hao Bai-Lin, {\it Chaos} (World Scientific, Singapore, 1984)}.
\item[{[8]}]{P. Cvitanovic, {\it Universality in Chaos} (Hilger, Bristol,
1984);
F. C. Moon, {\it Chaotic and Fractal Dynamics} (Ichaca, New York, 1992).}
\item[{[9]}]{ P. Pryant and K. Wiesenfeld,
Phys. Rev. A  {\bf 33}, 2525 (1986);
S. T. Vohra, L. Fabiny, and K. Wiesenfeld,
Phys. Rev. Lett.  {\bf 72}, 1333(1994)}.
\item[{[10]}]{ N. Platt, E. A. Spiegel, and C. Tresser,
Phys. Rev. Lett.  {\bf 70}, 279 (1993);
P. W. Hammer {\it et. al.},
Phys. Rev. Lett.  {\bf 73}, 1095 (1994)}.
\item[{[11]}]{ Mingzhou Ding, E. Ott, and C. Grebogi,
Phys. Rev. E  {\bf 50}, 4228 (1994).}
\item[{[12]}]{ C. Grebogi, E. Ott, and J.A. York,
Phys. Rev. Lett.  {\bf 48}, 1507 (1982);
Phisica,  {\bf 7}D, 181 (1983)}.
\item[{[13]}]{ R. B. May,  Nature  {\bf 261}, 459 (1976);
 L. P. Kadanoff, Physics Today (Dec.) {\bf 36}, 46 (1983)}.
\item[{[14] }]{B.-L. Hao,\  Phys. Lett. A \ {\bf 86}, 267 (1981);
M. Franszek and P. Pieranski, Canadian J. Phys., {\bf 63}, 488(1985)}.
\item[{[15]}]{ J. M. Carlson, J. S. Langer, and B. E. Shaw,
 Rev. mod. Phys. {\bf 53}, 657(1994)}.
\item[{[16]}]{ Q.-L. Li, L. Yu, B.-L. Hao, and J.-B. Chen, {\it Time and Space
Scanning of the Frequency-Magnitude Relation for Earthquakes}
(Chinese Earthquake Press, Beijing, 1979). }
\item[{[17]}]{ J.D. Farmer and J.J. Sidorowich, Phys. Rev. Lett. {\bf 59},
845(1987);M. Casdagli, Physica {bf 35}D, 335(1989);
H.D.I. Abarbanel, R. Brown and J.B. Kadtke, Phys. Rev. A {\bf 41}, 1782(1990);
J. Stark, J. Nonlinear Science, {\bf 3}, 197(1993)}.
\item[{[18]}]{ H.-P. Fang and L.-Y. Cao,
Phys. Rev. E  {\bf 51}, 6254 (1995)}
\item[{[19]}]{ L.-Y. Cao, H.-P. Fang, Q.-L. Li, and J.-B. Chen,
Int. J. Bif. $\&$ Chaos, accepted .
In this paper, the $b$-value for the seismic events in Tangshan region
in China is shown short-term predictable.}
\item[{[20]}]{ H.-P. Fang, G.W. He and L.Y. Cao,
Suppressing chaos and  transient
in parameter perturbed systems, Commun. Theor. Physics, accepted}.
\item[{[21]}]{ N. Packard, J. crutchfield, J.D. Farmer, and R. Shaw,
Phys. Rev. Lett.  {\bf 45}, 712(1980)}.
\item[{[22]}]{ C. Grebogi, E. Ott, and J.A. Yorke, Phys. Rev. Lett.
 {\bf 64}, 1196 (1990)}.

\end{description}
\section*{}
{\bf Figure Captions}
\begin{description}
\item[]{ Fig. 1\ The developing diagram of the logistic map in the parameter
interval $\mu \in [1.6,1.82]$ with a parameter step $\delta \mu = 10^{-5}$. The
initial point $x_0$ at $\mu =1.6$ is on the attractor.}
\item[]{ Fig. 2\ Same as Fig. 1 with
 $\mu \in [1.999, 2.003]$.}
\item[]{ Fig. 3\ The sliding average values of the Fig. 1 with
500 data a moving.}
\end{description}
\end{document}